\documentclass[aps,prl,amsmath,amssymb,twocolumn,showpacs,superscriptaddress]{revtex4}
\usepackage{graphicx}

\begin{document}

\author{E.Zaccarelli}
\affiliation{Dipartimento di Fisica and CNR-INFM-SOFT, Universita' di Roma La Sapienza, Piazzale Aldo Moro 2, 00185, Roma, Italy}
\author{C.Valeriani}
\author{E.Sanz}
\author{W. C. K. Poon}
\author{M. E. Cates}
\author{P. N. Pusey}
\affiliation{SUPA, School of Physics and Astronomy, The University of Edinburgh, James Clerk Maxwell Building, The King's Buildings,  Mayfield Road, Edinburgh, EH9 3JZ, UK }

\title{Crystallization of hard-sphere glasses}

\date{\today}

\begin{abstract}

We study by molecular dynamics the interplay between arrest and crystallization in hard spheres. For state-points in the plane of volume fraction ($0.54\le\phi\le0.63$) and polydispersity ($0\le s\le 0.085$), we delineate states that spontaneously crystallize from those that do not.
For non-crystallising (or pre-crystallization) samples we find  
isodiffusivity lines consistent with an ideal glass transition at $\phi_g \approx 0.585$, independent of $s$. Despite this, for $s<0.05$, crystallization occurs at $\phi > \phi_g$. This happens on timescales for which the system is ageing, 
and a diffusive regime in the mean square displacement is not reached; by those criteria, the system is a glass. 
Hence, contrary to a widespread assumption in the colloid literature, occurrence of spontaneous crystallization within a bulk amorphous state does not prove that this state was an ergodic fluid rather than a glass.

\end{abstract}


\maketitle



Pioneering computer simulations  in the 1950s
predicted that a system of monodisperse hard spheres (HS) should crystallize at high enough volume fraction, $\phi$ \cite{alder}. 
Thirty years later,  
a full phase diagram (from fluid densities up to random close packing, $\phi_{RCP} \approx 0.64$) was measured in a suspension of hard-sphere PMMA 
colloids \cite{pusey1986}. Alongside the fluid and crystal phases, this reported the formation of a glass at $\phi\ge\phi_g \approx 0.58$ \cite{pusey1986,pusey2_PRL_1987}. The existence and nature of the HS glass remains controversial \cite{vanmegen1993,vanmegenPRE1998,elmasri2005,vanmegenPRL2008}: 
for instance, a recent report ~\cite{brambilla} suggests that HS suspensions remain ergodic well beyond $\phi=0.58$. 
Ergodicity depends on observation time, so this finding does not rule out colloidal glasses on timescales of several hours.
(It does rule out an {\em ideal} glass transition, in which the structural relaxation time for $\phi>\phi_g$ is strictly infinite, as predicted by mode-coupling theory (MCT) \cite{MCT}.) 
For many years, a number of colloid physicists have held that, because particles in a glass cannot rearrange diffusively, crystallization within a homogeneous bulk glass cannot proceed unless it is seeded with pre-formed nuclei \cite{footone}. 
This view is stated explicitly in \cite{vanmegen1993} (but see \cite{noretract}); in combination with the observation that hard sphere colloids at $\phi > \phi_g$ show rapid bulk crystallization in microgravity, it has led Chaikin and others to assert that such colloids do not have a glass transition at all \cite{gravity}.

An important factor in studies of all these phenomena is polydispersity. Fractional standard deviations in particle size, $s$, of a few percent are inevitable experimentally. Some authors hold that polydispersity is essential to the formation of a hard sphere glass \cite{kawasaki95,vanmegenPRE2001} because a putative monodisperse sample would crystallize too 
rapidly \cite{rintoul}. (Here, as for glasses generally, the contest between vitrification and crystallization is usually portrayed as happening {\it during} rather than {\em after} a quench \cite{angelnature}.) In colloids, polydispersity
could play two distinct roles. First, it 
might influence cage formation, shifting any glass transition point. 
However, computational \cite{FoffiPRL2003_91_085701,SearJCP2000_113_4732} and experimental \cite{henderson1996} 
work seems to suggest that modest polydispersity ($s\le 0.1$) has little effect on structural relaxation times.   
On the other hand, numerical \cite{kawasaki95,auer2001} and experimental~\cite{henderson1996,vanmegenJCP2007} studies have shown that crystallization is strongly suppressed even at small finite $s$.
Hence a second possible role of polydispersity is to  destabilize directly the crystal. 
Indeed work on the equilibrium HS phase diagram \cite{barrat,kofke,bartlett,sollich} predicts that beyond a `terminal' $s$ value $(\sim 0.07$) there is no stable crystal of the same composition as the fluid. Crystallization then requires size fractionation \cite{sollich}, which involves transport over  large distances.

In this Letter we present molecular dynamics (MD) simulations that probe in  detail 
the effects of polydispersity on the HS glass transition and on crystallization. 
We find almost no effect of polydispersity on the dynamics of diffusive rearrangement within the amorphous state
near the glass transition, but confirm also that polydispersity strongly suppresses 
crystallization \cite{henderson1996,kawasaki95}.  More strikingly, we find that for low polydispersity ($s\le5\%$), 
hard spheres crystallization can occur spontaneously within a bulk amorphous state that is plainly a glass (no particle diffusion) on the timescale of observation (see Fig.~3 below). Note that, just as in colloids \cite{vanmegen1993,footone} non-diffusive pathways to crystal {\em growth} --from a pre-seeded nucleus, within a bulk glass-- are well-established for conventional (microscopic) glass-formers \cite{growth}. Experimental evidence for crystallization in unseeded homogeneous bulk glasses exists \cite{metallic}, but is much less clear: surface devitrification and residual diffusion are both hard to rule out. Our simulations establish that crystallization without diffusion is indeed possible in the simplest of all glasses: bulk homogeneous hard spheres. We plan further detailed analyses to elucidate the mechanism; future colloidal experiments could equally illuminate the microphysics involved. 

Our MD study addresses $N=2000$ hard spheres  
in an $NVT$ ensemble, with $0.54 \le \phi \le 0.63$ and $0\le s\le 0.085$, using an event-driven algorithm for particles interacting
via hard potentials.
To approximate a  bulk system, we use periodic boundary conditions 
in a cubic box (volume $V$). 
Mass, length and time are measured in units of the particle mass $m$, the mean particle diameter $\overline{\sigma}$, and the thermal timescale
$t_0=\bar\sigma (m/k_B T)^{1/2}$, where $k_BT=1$.  
Polydispersity is represented by a 31 component discrete Gaussian 
distribution of standard deviation $s$; the volume fraction obeys $\phi \equiv \pi N\overline{\sigma^3}/6V$.
At each `state point', defined by given values of $\phi$ and $s$, we run simulations of duration $t_{M}= 10^5$; to 
improve the statistics we perform 5 independent runs. Each run is started from a different initial 
configuration; those containing ordered seeds are eliminated by requiring an initially low percentage (less than $5\%$) of `crystalline particles'.
(We check that any remaining patches of initial order do not act as nuclei.)
Such particles are defined locally using a rotationally invariant bond order parameter $d_6$~\cite{tenwolde}, 
with cutoffs set at $r_c=1.4$
(to define each particle's first neighbors), $d_c=0.7$ (a first 
criterion that identifies candidate crystalline particles 
as having $d_6 \ge d_c$) and $N_c=6$ (a second and more stringent criterion requiring such a particle to have at least $N_c$ neighbors that also satisfy $d_6 \ge d_c$). For additional details and results of the simulations, see \cite{RSpeter}.

\begin{figure}[h!]
\includegraphics[width=8cm, clip=true]{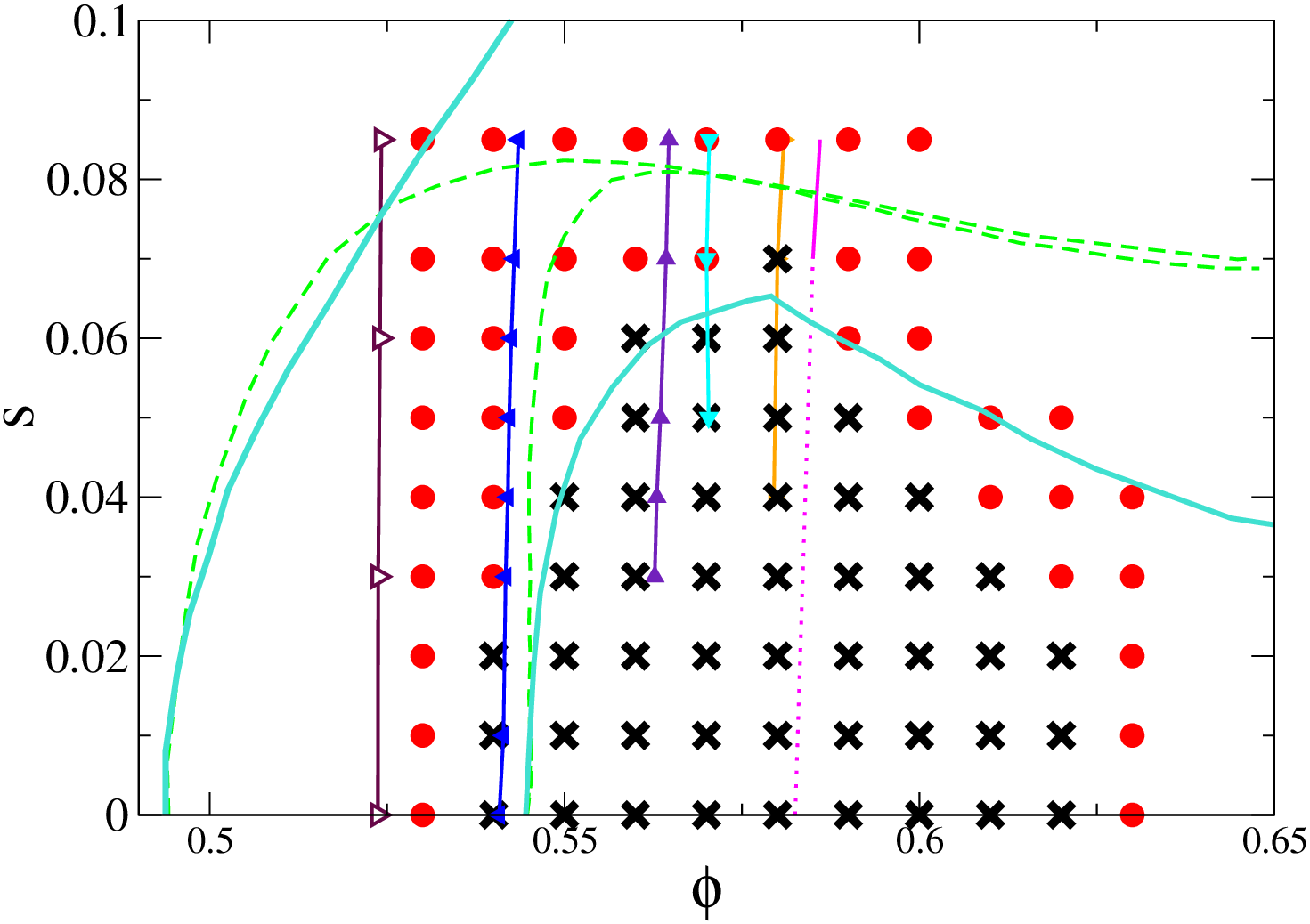}
\caption{Stability diagram in the $(\phi, s)$ plane, reporting state points that crystallize (black crosses) in at least one of five runs, and state points where all runs remain 
disordered until  $t_{M}$ (red circles). We also show as triangles 
isodiffusivity lines  (left to right: $D=5\times 10^{-3}$(blue), $D=10^{-3}$(violet), $D=5\times 10^{-4}$ (cyan), D=5 10$^{-5}$(orange)); 
to the right of these lie the MCT glass transition for a binary mixture of matched $s$ (maroon, open triangles),
 and the extrapolated ideal glass 
line from our data (magenta), with a dotted continuation into the crystal as guide to the eye. 
The phase boundaries allowing~\cite{bartlett} and not allowing~\cite{sollich} for size fractionation are shown 
as solid (turquoise) and dashed (green) lines respectively.} 
\label{fig:phase}
\end{figure}

For each state point we track the pressure, and the fraction of crystalline particles, over time: a state point is declared `crystalline' if at least one 
of the five simulation runs develops a proportion of crystalline particles exceeding 85\%, within $t_{M}$. 
The resulting stability diagram in the $(\phi ,s)$
plane is shown in Fig.~\ref{fig:phase}. This broadly agrees with, but vastly extends, the much earlier simulation results 
of \cite{kawasaki95}.

Fig.~\ref{fig:phase} shows that crystallization takes place 
in a well-defined region of the stability diagram. 
The maximum polydispersity observed, $s\simeq 0.07$, almost coincides with the predicted limit of existence of a non-fractionated solid \cite{bartlett}. Polydispersity is relevant on both the 
low and the high-$\phi$ borders of the crystallising region, narrowing
to a similar extent the crystallizing range from both sides.
The low-$\phi$ part of the boundary almost tracks the predicted density of the fluid at coexistence, with a fixed supersaturation $\Delta\phi \approx 0.05$. Thus the effect of $s$ on crystallization at low $\phi$ can be accounted for simply by the shift 
of thermodynamic equilibrium boundaries (which reduces supersaturation at a given $\phi$), combined with an assumption that a nucleation barrier is only surmountable for sufficient $\Delta\phi$ \cite{auer2001}. On the other hand, as stated previously, suppression of crystallization at high $\phi$ could stem from slowed kinetics in the amorphous phase, or direct destabilization of the crystal. To distinguish these alternatives, we next assess the impact that polydispersity has on structural relaxation dynamics within the amorphous state. 

\begin{figure}[h!]
\includegraphics[width=8cm, clip=true]{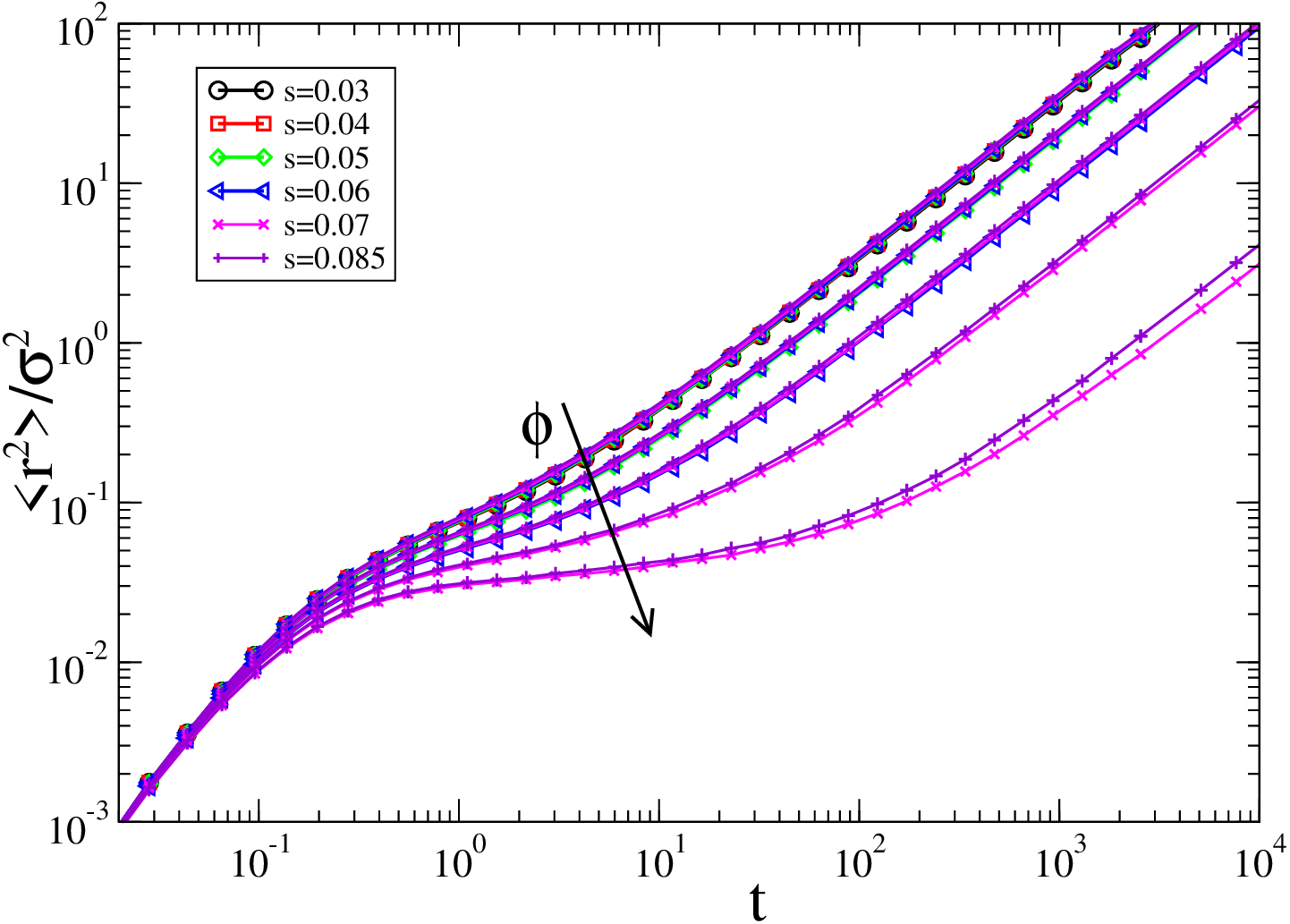}
\includegraphics[width=8cm, clip=true]{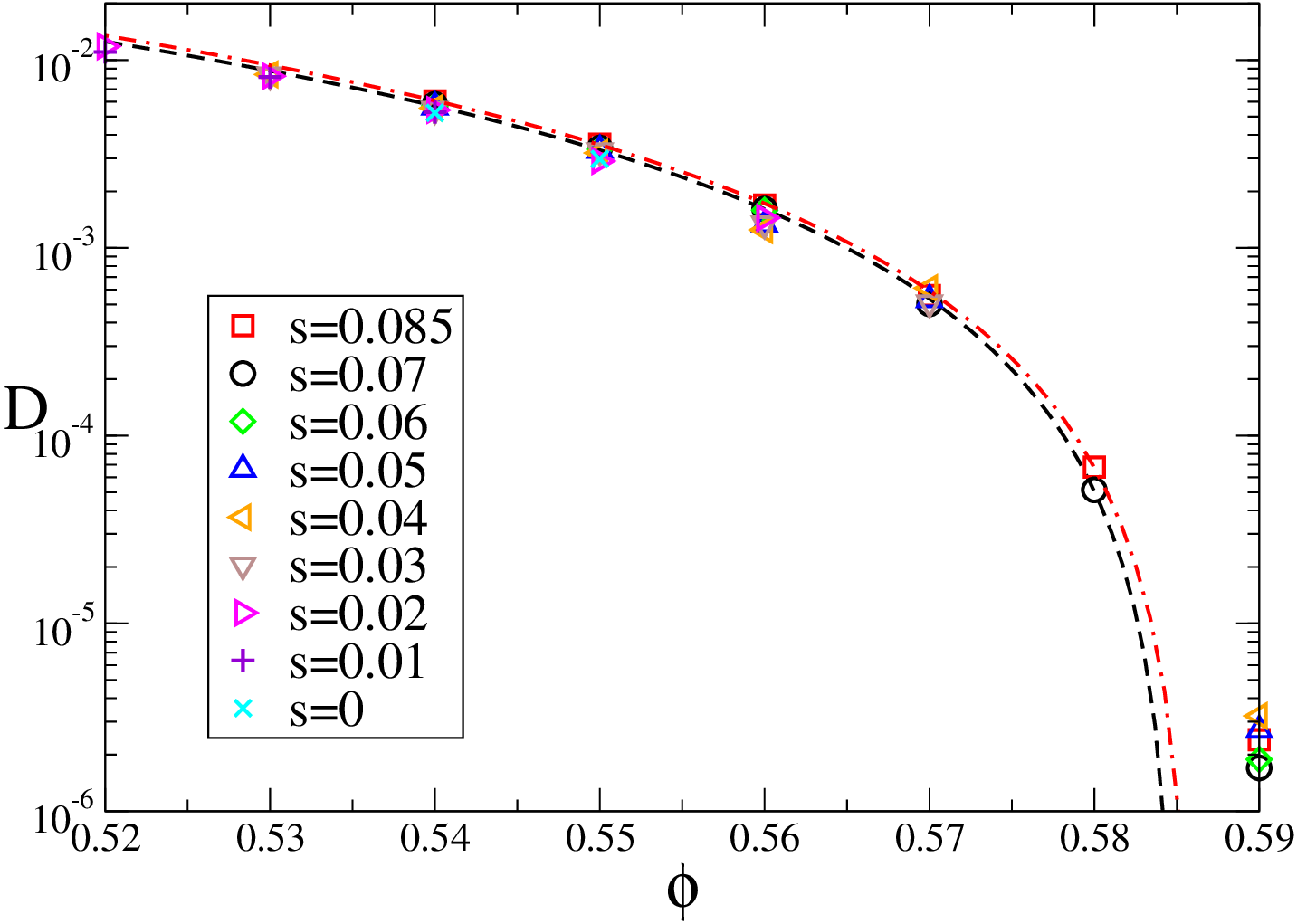}
\caption{(top) Time evolution of the average MSD for  
$\phi=0.54, 0.55, 0.56, 0.57, 0.58$ (from top left to bottom right) 
for all non-crystallising state points. 
Most symbols are indistinguishable due to the almost perfect collapse of 
data for different $s$. (bottom) Extracted $D$ values 
versus $\phi$ (symbols), with corresponding power-law fits for $s=0.07$ 
(dashed line) and for $s=0.085$ (dashed-dotted line). $D$ values at $\phi = 0.59$ show ageing (see below) and are excluded from the fit.}
\label{fig:MSD}
\end{figure}

For this we require quantitative data on long-time diffusion, and so must focus on those state points which remain disordered during our
simulation time. For each of these we calculate the mean square displacement (MSD) and 
extract from its long-time limit the self-diffusion coefficient $D$ (averaged over 5 runs). 
In Figure~\ref{fig:MSD}~(top) we plot the run-averaged MSD curves ($\langle r^2(t) \rangle$)
for non-crystallizing state points at all $s$ with $0.54 \le \phi \le 0.58$. Of these, only for $s=0.07$
and $s=0.085$ can we study the full $\phi$ range directly. But for no  $\phi$ is there evidence of significant $s$-dependence in the diffusive rearrangement kinetics. 
Instead we recognise at all $s$ the same short and intermediate time behavior, and observe the emergence of the same plateau, typical of systems close to dynamic arrest, for $\phi \gtrsim 0.57$.


Fig.~\ref{fig:MSD}~(bottom) plots the extracted self-diffusion coefficients as a function $\phi$. To estimate an ideal glass line, we apply an MCT-inspired fit
$D\sim (\phi - \phi_g)^{\gamma}$, for the two polydispersities at which crystallization never intervenes. The data for $\phi = 0.59$ are excluded because these samples exhibit ageing (we discuss ageing further below). 
The fits give $\gamma \simeq 2.21$ and $\phi_g\simeq 0.586$ (for $s=0.085$) and 
$\gamma \simeq 2.15$ and $\phi_g\simeq 0.585$ (for $s=0.07$). Both fits are included in 
Fig.~\ref{fig:MSD}~(bottom); their difference is barely significant.
Using the computed $D$ values, we next estimate isodiffusivity lines
in the $\phi,s$ plane. These lines can be measured outside the 
crystallization region, and also some way into its interior 
(in cases where samples have exited the MSD plateau before they crystallize). 
The isodiffusivity lines shown in Fig.~\ref{fig:phase} are parallel as for calculations for other repulsive 
systems \cite{zacca2002,zacca-att2}, and almost vertical ($s$-independent). 
We also show in Fig.~\ref{fig:phase} the glass line from an MCT calculation 
for a binary mixture with $s$ at each point equal to that of our simulations \cite{voigtmann}, using as input the partial static structure factors
 calculated from additional simulations. On moving away from the monodisperse limit ($\phi_{MCT}\simeq 0.525$~\cite{MCT}),
this is near-vertical and tracks the isodiffusivity and extrapolated ideal glass lines.


The above data presents strong evidence that polydispersity up to 0.08 has barely any effect on the dynamics of structural rearrangement in amorphous HS systems on the approach to arrest; independently of polydispersity, these show a glass transition (ideal or otherwise) at
$\phi = \phi_g \approx 0.585$. This points firmly toward destabilization of the crystal as the reason for polydispersity to suppress crystallization.  Remarkably though, Fig.~\ref{fig:phase} also shows that systems of low polydispersity ($s \le 0.05$) can crystallize for $\phi$ well beyond the extrapolated $\phi_g$. 

\begin{figure}[h!]
\includegraphics[width=8cm, clip=true]{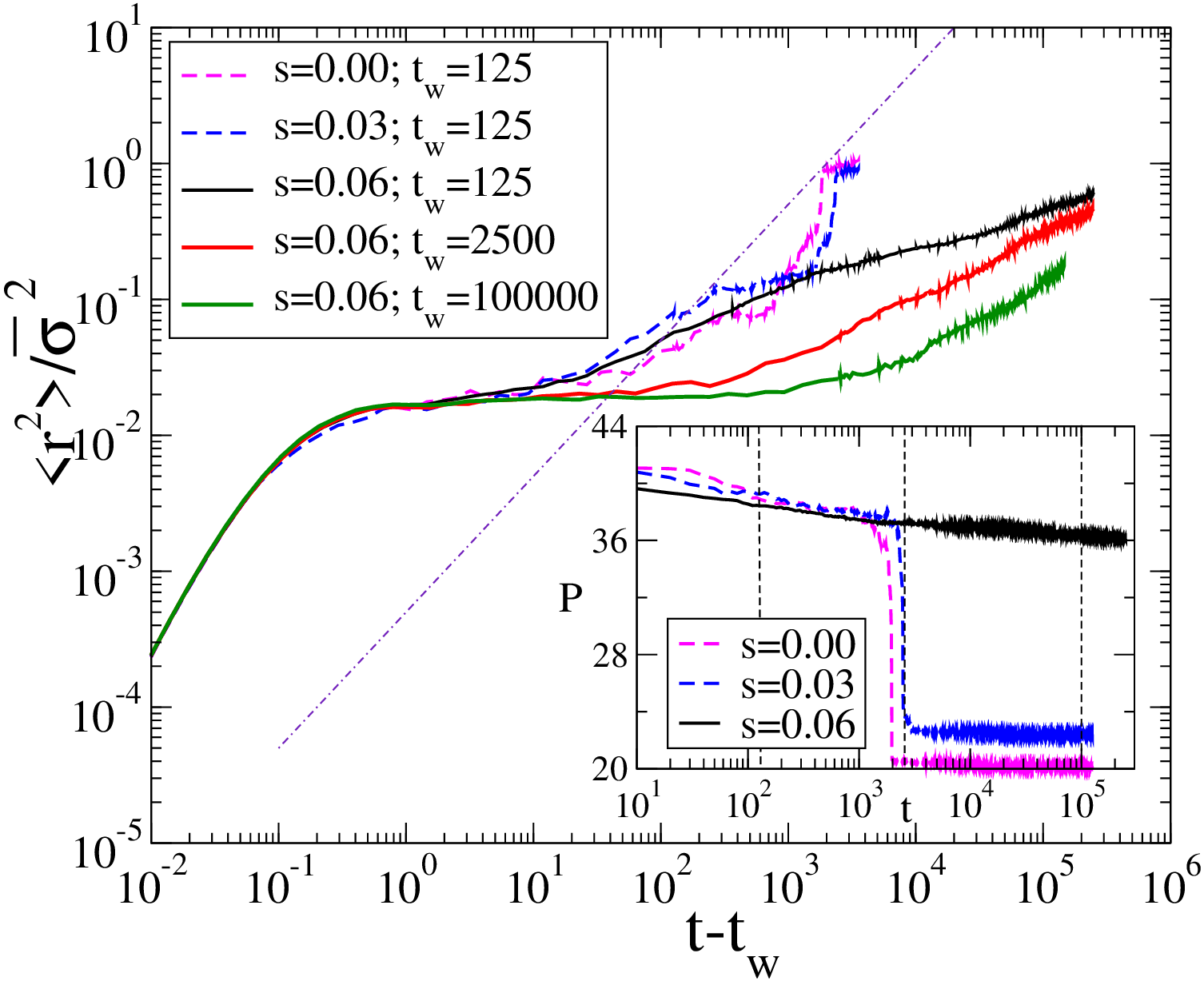}
\caption{
MSD versus $t-t_w$ for $\phi=0.60$ and $s=0.06$, at various $t_w$, averaged over 9 different runs. 
The MSD for the same $\phi$ and lower $s$ $(s = 0$ and $0.03)$ is also included. At these low $s$ 
the system crystallizes, and averaging over different runs is no longer possible. As a consequence, the MSD is
more noisy but, within the statistical error, identical to that of $s=0.06$ until crystallization takes place. 
The dashed-dotted line, with slope 1, has been included to stress the fact that the system is subdiffusive
at the moment it crystallizes. 
Inset: Pressure versus time for the same state
points. Vertical lines indicate the $t_w$-values employed in the MSD
calculations.}
\label{fig:aging}
\end{figure}

The most obvious explanation for this would be that these systems are after all ergodic: the glass transition is not ideal. What matters though is whether they are ergodic on the simulation timescale.
To check this we have studied closely the case of $\phi = 0.6$, 
for both 
crystallizing ($s \le 0.03$) and non-crystallizing ($s > 0.03$) polydispersity. 
In Figure~\ref{fig:aging} we report the MSD as a function of $t-t_w$ 
for $\phi=0.60$ and $s=0.06$, 
where $t_w$ is the ageing time elapsed from the beginning of the run.
While the short-time and
intermediate plateau regimes remain essentially unchanged, a
significant ageing of the dynamics appears at long times with increasing
$t_w$. In the inset, we report the pressure versus time 
for the same state points. For $\phi = 0.60$ and $s=0.06$ 
the pressure does not equilibrate to a clear plateau, but continuously decreases: 
hence this system is nonergodic on all time scales simulated ($t<t_{M} = 10^5$). Invoking now our previous evidence that polydispersity has no discernible effect on dynamics in the amorphous state, we can infer that the same nonergodicity is present in amorphous states at $\phi = 0.60$ and $s \le 0.03$, which do crystallize. 
Indeed, by monitoring the MSD in the MD simulations, we find that these near-monodisperse samples crystallize on 
time scales $\sim~10^3$ (shown in Fig.~3 by a sharp rise of the MSD and a sharp pressure drop), at least two orders
of magnitude smaller than the timescale to attain ergodicity (inferred from the ageing studies). 
Fig.~3 confirms that the MSD curve found (on a fresh sample) at high $s$
is closely tracked by the corresponding low-$s$ samples, 
until interrupted by crystallization.
Moreover, on these crystallization time scales, particles have diffused, on average, only a few tenths of a diameter.

We conclude that dense hard spheres can crystallize on a time scale short compared to that of particle diffusion in the initially homogeneous, amorphous state. Put differently, bulk hard sphere {\em glasses} can crystallize. The mechanism of this transition, direct from amorphous solid to crystal, is currently under study.
A strong possibility, already proposed in \cite{vanmegenPRE2001}, is
that this direct crystallization channel is closed only when the terminal polydispersity is reached. Only beyond this point is diffusion needed, to allow fractionation to occur. (Suggestively, the crystallizing region in Fig.~\ref{fig:phase} lies wholly within the unfractionated crystal zone on the equilibrium phase diagram of \cite{bartlett,sollich}.) Note also that, to equate pressure, the crystal is denser than the glass from which it forms \cite{footone}. 

Our findings undermine an implicit argument in the colloid literature \cite{pusey1986,vanmegen1993,gravity} that if an amorphous state readily forms crystals in bulk, this state must have been an ergodic fluid, not a glass. 
This now appears unsustainable, and one can no longer infer \cite{gravity}, from the observed rapid crystallization 
at $\phi>\phi_g$ of colloids in microgravity, that such colloids have no glass transition. 
The view that crystallizaton implies mobility is anyway rather at odds with another longstanding view, 
that truly monodisperse systems crystallize on a timescale so fast that a glass cannot be seen 
(see \cite{kawasaki95}  for a discussion). This view also appears overstated; our results for $\phi = 0.63$ and $s=0$ show glassy dynamics without crystallization up to $t \simeq 10^5$ (an experimentally accessible range for colloids).

In this Letter we have studied by simulation the interplay between dynamics and crystallization in both monodisperse and polydisperse hard spheres.
We have computed a stability diagram in the volume fraction / polydispersity plane, showing the region where crystallization 
takes place. 
This region is controlled at low $\phi$
by supersaturation (the thermodynamic driving force for crystal nucleation) and at high $\phi$
by an interplay between polydispersity and dynamical arrest.  
Computing the mean square displacement, we concluded that 
polydispersity barely affects the dynamics of diffusive rearrangement in the amorphous state. 
An MCT extrapolation of the isodiffusivity lines 
yields,  for all studied $s$ (including $s=0$), 
an ideal glass transition point at $\phi_g \approx 0.585$. 
For $\phi>\phi_g$, the system is nonergodic, and shows ageing. 
However for near-monodisperse samples this process is interrupted by a glass-to-crystal transition, 
whose mechanism we hope to elucidate in a future work. We conclude, crucially for colloid experiments, 
that formation of a crystal does not prove absence of a glass. 

We thank G. Foffi for his contributions to the early stages of this work. This work was funded by UK EPSRC (EP/E030173/1, EP/D071070/1), the 
EU  MRTN-CT-2003-504712, and NMP3-CT-2004-502235 (SoftComp). MEC is supported 
by the Royal Society. Computer resource was provided by ECDF which is 
partially supported by eDIKT.


\bibliographystyle{./apsrev}
\bibliography{./HSpaper17}

\end{document}